\documentclass[%
 reprint,
 amsmath,amssymb,
 aps,
]{revtex4-2}

\usepackage{graphicx}
\usepackage{dcolumn}
\usepackage{bm}
\usepackage{xcolor}

\begin{document}
\newcommand{\red}[1]{{\color{red} #1}}
\newcommand{\blue}[1]{{\color{blue} #1}}


\title{Exact analytic photon escape cones and silhouettes for extremal Kerr}

\author{Joshua Baines}
 \email{joshua.baines@sms.vuw.ac.nz}
\author{Matt Visser}%
 \email{matt.visser@sms.vuw.ac.nz}
\affiliation{%
School of Mathematics and Statistics, Victoria University of Wellington, Wellington 6140, New Zealand.
}%

\date{Wednesday 15 May 2024; \LaTeX-ed \today }

\begin{abstract}
We exhibit exact analytic, fully explicit, photon escape cones and related silhouettes (``shadows'') at arbitrary distance from  the extremal Kerr black hole. In contrast to the situation for non-extremal Kerr, where one typically has to resort to either numerical and/or perturbative investigations, the situation for extremal Kerr is amenable to direct analytic investigation. 
\end{abstract}

\maketitle


\def\rg{{r_\gamma}}
\def\rp{{r_\gamma}}
\def\ro{{r_*}}
\def\O{{\mathcal{O}}}

\emph{Introduction:} Calculations and images of silhouettes (``shadows'') cast by various black holes have recently attracted much interest, both for theoretical~\cite{Cunha:2018, Gralla:2019,  Bambi:2019,  Abdujabbarov:2016, Hioki:2009, Johannsen:2010,  Broderick:2013, Broderick:2008, Johannsen:2015, Cardoso:2016, Claudel:2000, Tsukamoto:2014, Tsukamoto:2017, Ogasawara:2019mir, Ogasawara:2020frt, Zulianello:2020cmx} and observational~\cite{EHT:2019a, EHT:2019d, EHT:2019e, EHT:2019f, EHT:2022a, EHT:2022e, EHT:2022c, EHT:2022f, Vagnozzi:2022} reasons. These silhouettes are just the time reversed complement of the escape cones of photons emitted from regions near the horizon, a topic which has its own even longer history.

In spherical symmetry  a lot can be done analytically, ranging from early work by Synge in the mid 1969's~\cite{Synge:1966} to recent 
work by the current authors~\cite{Baines:2023}.
However, once one adds rotation the situation is considerably messier. While a number of limited (but useful) analytic results are known, for specific calculations one often has to resort to numerical/\-graphical techniques, or resort to (first and second order) perturbative techniques~\cite{Baines:2024}.

One special case that is (perhaps surprisingly) amenable to full analytic  analysis (albeit with interesting subtleties) is extremal (maximally rotating) Kerr, (where $a\to m$), a situation which we shall fully explore in the current article. 

\emph{Key result:}
Let the coordinates $(\theta_*,r_*)$ denote the declination and radial position of the emission point, and the coordinates $(\Theta,\Phi)$ denote the edge of the escape cone on the celestial sphere of the emission point.
Equivalently, let the coordinates $(\theta_*,r_*)$ denote the declination and radial position of the observer, and the coordinates $(\Theta,-\Phi)$ denote the edge of the silhouette on the celestial sphere of the observer.
Define
 the two dimensionless quantities
\begin{equation}
\label{E:w}
w(\theta_*,\Phi) =\sin\theta_*\sin\Phi+ \sqrt{[1+\sin\theta_*\sin\Phi]^2 +\cos^2\theta_*},
\end{equation}
and $z = m/r_* \in (0,1)$.
Then:\\
Case I: In the region where $w(\theta_*,\Phi)\geq 0$ we have
\begin{equation}
\label{E:sin_wz}
\sin\Theta = {2 (1+w) \, z \, (1-z)\over1+(w^2-1) z^2}.
\end{equation}
Case II: In the region where $w(\theta_*,\Phi)< 0$, implying
\begin{equation}
\sin\theta_* > \sqrt{3}-1, 
\quad\hbox{and}\quad
\sin\Phi < -{1+\cos^2\theta_*\over 2\sin\theta_*},
\end{equation}
we have
\begin{equation}
\label{E:sin_wz3}
\sin\Theta = 
- {z\over 1+z} \;{1+\cos^2\theta_*\over \sin\theta_*} \;{1\over\sin\Phi}.
\end{equation}

These two cases completely characterize the escape cone (and, equivalently,  capture cone and  silhouette). 
At the boundary between these two regions,  we have both $\sin\theta_*>  \sqrt{3}-1$  and $\sin\Phi = -(1+\cos^2\theta_*)/(2\sin\theta_*)$, so $w\to 0$, and the two results converge on each other:
\begin{equation}
\label{E:sin_wz2b}
\sin\Theta \to {2  \, z \over1+ z}.
\end{equation}
Note that for $\sin\theta_*< \sqrt{3}-1$, sufficiently close to the axis of rotation, the ``anomalous'' $w<0$ region is empty and one only has to deal with Case I.
Somewhat related comments can be found scattered throughout  many parts of the literature, though often presented in rather different forms.  
For comparison purposes it is useful to note
\begin{equation}
\arcsin(\sqrt{3}-1) = 0.8213274620= 47.05859716^\circ
\end{equation}
and 
\begin{equation}
\arcsin(\sqrt{3}-1) = \arccos \sqrt{2\sqrt{3}-3} = \arctan([4/3]^{1/4}).
\end{equation}

\emph{Known results for generic Kerr:}
The Kerr spacetime is extensively discussed in references~\cite{Kerr1,Kerr2,Kerr3,Kerr4, Kerr5,Kerr6,Kerr7,Kerr8,Kerr9}.
For black hole silhouettes
Perlick and Tsupko provide an extensive review  in reference~\cite{Perlick:2021}, specifically addressing the Kerr spacetime. 
Two key equations determining the edge of the Kerr photon escape cone are~\cite{Perlick:2021, Tsupko:2017, Grenzebach:2014}:
\begin{equation}\label{sinTheta_general}
\sin\Theta =  {2\rg\sqrt{\rg^2-2m\rg+a^2} \sqrt{\ro^2-2m\ro+a^2} 
\over \ro^2(\rg-m) +\rp( \rp^2 -3 \rp m+2a^2)};
\end{equation} 
\begin{equation}\label{sinPhi_general}
\sin\Phi = {\rg^2(\rg-3m)+ a^2[\rg  + m + (\rg-m)\sin^2\theta_* ]
\over 2a\rg \sin\theta_* \sqrt{\rg^2-2m\rg+a^2}}.
\end{equation}
Here $r_\gamma$ denotes the  (\emph{a priori} unknown) location  of the ``spherical photon orbits'', constant-$r$ orbits in the Boyer--Lindquist $r$ coordinate. They sweep out segments of the constant $r$ \emph{topological} 2-sphere. (This is not \emph{geometrically} a constant-curvature 2-sphere.) Except for polar orbits where the entire 2-sphere is swept out, the ``spherical photon orbits'' sweep out an equatorial zone of the topological 2-sphere. Another special case are the equatorial orbits --- where spherical orbits reduce to circular orbits.
These expressions are  subject to the physical constraint that the spherical photon orbits must lie outside the (outer) horizon $\rg > r_H=m+\sqrt{m^2-a^2}$. 

\emph{Proof of Case I:} As long as one enforces $\rg>r_H$ one can safely take the limit $a\to m$ in equations (\ref{sinTheta_general}) and (\ref{sinPhi_general}) to obtain
\begin{equation}\label{sinTheta_supra}
\sin\Theta =  {2\rg (r_*-m) 
\over  \rp^2 -2 m \rp +r_*^2)};
\end{equation} 
\begin{equation}\label{sinPhi_supra}
\sin\Phi = {\rg^2-2m\rg- m^2 \cos^2\theta_* 
\over 2m\rg \sin\theta_*}.
\end{equation}
It is convenient to define $\rg = m(1+w)$ and $z = m/r_*$.
Then equations (\ref{sinTheta_supra}) and (\ref{sinPhi_supra}) become
\begin{equation}\label{sinTheta_supra2}
\sin\Theta =  {2(1+w)z(1-z)
\over  1+ (w^2-1) z^2};
\end{equation} 
\begin{equation}\label{sinPhi_supra2}
\sin\Phi = - {2+w^2 - \sin^2\theta_* 
\over 2(1+w)\sin\theta_*}.
\end{equation}
Solving (\ref{sinPhi_supra2}) for $w$ leads to equation (\ref{E:w}) as claimed. 
This completes the derivation of Case I. 

\emph{Proof of Case II:} What instead happens when the naive formula for $w$ gives unphysical negative answers? In this case the extremal limit must be considered with more subtlety. 
Build the constraint $\rg > r_H >m$ into the formalism before taking the extremal limit, but do so in a way that permits $\rg \to m$. 
First define $a=m\sqrt{1-\epsilon^2}$ with $\epsilon\in[0,1]$ so that 
\begin{equation}
r_H=m+\sqrt{m^2-a^2} = m(1+\epsilon). 
\end{equation}
Now define
\begin{equation}
\rg = m(1 + \epsilon \cosh \zeta); \qquad \zeta\in[0,\infty].
\end{equation}

On the one hand this is merely a definition. On the other hand, keeping $\epsilon\in[0,1]$ and $\zeta\in[0,\infty]$ in their respective physical ranges will in turn force $\rg$ to stay in the physical regime $\rg\geq r_H$, while simultaneously enforcing $\rg\to r_H \to m$ in the extremal limit. 
We also set $z=r_H/r_*= m(1+\epsilon)/r_* \in (0,1]$.

\begin{widetext}
With these definitions we have the (still exact) results
\begin{equation}
\sin\Theta = 
{2z\sqrt{1-z} \sqrt{(1+\epsilon)^2-z(1-\epsilon^2)} 
(1+\epsilon\cosh\zeta) \sinh\zeta
\over
((1+\epsilon)^2 - (1+2\epsilon^2)z^2)\cosh\zeta -2\epsilon z^2 +\epsilon^2 z^2 \cosh^3\zeta}
\end{equation}
and 
\begin{equation}
\sin\Phi = - {(1+2\epsilon^2 + (1-\epsilon^2) \cos^2\theta_* ) \cosh\zeta + 2\epsilon -\epsilon^2\cosh^3\zeta
\over
2\sqrt{1-\epsilon^2} (1+\epsilon\cosh\zeta) \sin\theta_* \sinh\zeta}. 
\end{equation}
\end{widetext}

The extremal limit $\epsilon\to0$ is now particularly simple
\begin{equation}
\sin\Theta = {2z\over 1+z} \tanh \zeta,
\end{equation}
and 
\begin{equation}
\sin\Phi = -{1+\cos^2\theta_*\over 2\sin\theta_*} \coth \zeta.
\end{equation}
Note that this last limit only makes sense for the region
\begin{equation}
\label{E:near-horizon-limit}
\sin\Phi \leq  -{1+\cos^2\theta_*\over 2\sin\theta_*},
\end{equation}
which corresponds to $w\leq0$;  a region which is now seen to correspond to a careful implementation of the near-horizon limit. Eliminating $\zeta$ in this near-horizon limit we see
\begin{equation}
\sin\Theta = 
- {z\over 1+z} \;{1+\cos^2\theta_*\over \sin\theta_*} \;{1\over\sin\Phi}.
\end{equation}
This completes the derivation of Case II. 

\emph{Comments on Case II:}
Over the region of interest, specified by  equation (\ref{E:near-horizon-limit}),  this result for $\sin\Theta$ never diverges, and we can in fact write
\begin{equation}
\sin\Theta = 
 {z\over 1+z} \;{1+\cos^2\theta_*\over \sin\theta_*} \;
 {1\over|\sin\Phi|}.
\end{equation}
At the edge of this near-horizon regime one has
\begin{equation}
\sin\Theta \to {2z\over 1+z}; 
\qquad
\sin\Phi \to  -{1+\cos^2\theta_*\over 2\sin\theta_*}.
\end{equation}
It is sometimes worthwhile to define
\begin{equation}
\sin\Phi_{edge} = -{1+\cos^2\theta_*\over 2\sin\theta_*},
\end{equation}
and in the Case II region write 
\begin{equation}
\label{E:edge-Theta}
\sin\Theta = 
 {2z\over 1+z} \;{\sin\Phi_{edge}\over\sin\Phi} = 
 {2z\over 1+z} \;{|\sin\Phi_{edge}|\over|\sin\Phi|} .
\end{equation}
One can also re-cast the discussion for Case II in terms of Cartesian coordinates $(x,y)=(\sin\Theta\cos\Phi, \sin\Theta\sin\Phi)$.

In these Cartesian coordinates
\begin{equation}
y =  - {2z\over 1+z} \;|\sin\Phi_{edge}|
\end{equation}
and
\begin{equation}
x 
= y \cot\Phi
= - {2z\over 1+z} \; |\sin\Phi_{edge}|\; \cot\Phi.
\end{equation}
This makes it clear that in Cartesian coordinates one is dealing with a (constant-$y$) straight line in the $x$-$y$ plane.

\emph{Discussion:}
As we have seen, while exact analytic results are easily extracted for exremal Kerr, there are still some subtleties involved in the analysis: photon emission from the axis of rotation, or from the region sufficiently near the axis of rotation, is relatively simple (Case I). Photon emission from anywhere on the equatorial slice, or from a precisely defined region near the equatorial slice, requires some delicate analysis concerning the range of validity of certain intermediate formulae (Case II).

A nice feature of the anlaysis is that we have not yet had to place any \emph{a priori} estimate of $r_*$ (apart from the minimalist physical constraint $r_* > m$) so that the same analysis applies simultaneously to both silhouettes and escape cones (and one does not have to be in any asymptotic regime). 

\emph{Large-distance silhouettes:} If one wishes specifically to focus on galactic black hole silhouettes then one is certainly working in the regime $z=m/r_* \ll 1$ and the discussion simplifies somewhat.\\
Case I: In the region where $w(\theta_*,\Phi)\geq 0$ we have
\begin{equation}
\label{E:sin_wz_sh1}
\sin\Theta = 2 (1+w) \, z \, +\O(z^2).
\end{equation}
(With $w$ as in equation (\ref{E:w}).)\\
Case II: In the region where $w(\theta_*,\Phi)< 0$, we have
\begin{equation}
\label{E:sin_wz_sh2}
\sin\Theta = 
- {z} \;{1+\cos^2\theta_*\over \sin\theta_*} \;{1\over\sin\Phi}
+\O(z^2).
\end{equation}
That is
\begin{equation}
\label{E:sin_wz_sh2}
\sin\Theta = 
 2z \;{\sin\Phi_{edge}\over\sin\Phi}
+\O(z^2).
\end{equation}
In this large-distance silhouette limit (but not in general) both Cases I and II are linear in $z$.

\emph{Near-horizon escape cones:}
If instead one wishes to focus on near-extremal-horizon black hole escape cones then one is working in the regime $z=m/r_* \approx 1$ and (modulo some subtleties) the discussion again  simplifies somewhat.\\
Case I: 
In the region where $w(\theta_*,\Phi)\geq 0$,  keeping $w$ fixed,  (so keeping $(\theta_*,\Phi)$ fixed), we have
\begin{equation}
\label{E:sin_wz_sh1}
\sin\Theta = {2 (1+w)(1-z)\over w^2}  \, +\O([1-z]^2).
\end{equation}
So keeping $w>0$ fixed the escape cones narrow to zero as the emission point approaches the horizon.

\newpage\noindent
If instead we consider the physically different situation where we keep $z$ fixed,  and let $(\theta_*,\Phi)$ move to the edge of the Case I  region, then we have
\begin{equation}
\label{E:sin_wz_sh1}
\sin\Theta = {2 z\over 1+z}  \, +\O(w).
\end{equation}
Then exactly on the edge of the Case I region ($w=0$)
\begin{equation}
\label{E:sin_wz_sh1}
\sin\Theta = {2 z\over 1+z}.
\end{equation}
So, keeping $w=0$ fixed, $\sin\Theta \to 1$  as the emission point approaches the horizon.\\
Case II: In the region where $w(\theta_*,\Phi)< 0$, we have
\begin{equation}
\label{E:sin_wz_sh2}
\sin\Theta = 
- {1\over2} \;{1+\cos^2\theta_*\over \sin\theta_*} \;{1\over\sin\Phi}
+\O(1-z).
\end{equation}
That is
\begin{equation}
\label{E:sin_wz_sh2}
\sin\Theta = 
 \;{\sin\Phi_{edge}\over\sin\Phi}
+\O(1-z).
\end{equation}
At the edge of the Case II region $\sin\Theta\to 1+\O(1-z)$, and, if the emission point subsequently approaches the horizon, $\sin\Theta\to 1$.  

\emph{Intermediate region:} To drive home the fact that our analysis is not limited to  asymptotic regions, consider $r_*=3m$, that is $z=1/3$. This corresponds to a source or observer well outside the horizon but still deep in the gravitational field. \\
Case I:
In the region where $w(\theta_*,\Phi)\geq 0$ we have
\begin{equation}
\label{E:sin_wz_sh1}
\sin\Theta = {4 (1+w)\over 8+w^2} \in \left[{1\over2},1\right].
\end{equation}
Case II:
In the region where $w(\theta_*,\Phi)< 0$ we have
\begin{equation}
\label{E:sin_wz_sh2}
\sin\Theta = 
 {1\over2} \; {\sin\Phi_{edge}\over\sin\Phi} 
 \in \left[{1\over2}|\sin\Phi_{edge}|,{1\over2}\right].
\end{equation}

\emph{Conclusions:}
We have seen that for extremal Kerr spacetime the exact analytic \emph{shape} of the escape cones/silhouettes can reasonably easily be extracted from the general formalism. 
We had initially hoped to also be able to analytically calculate exact \emph{solid angles} for these escape cones/silhouettes, but the relevant integrals proved to be intractable except in very special cases. 
 
In the longer run it would also be interesting to perturbatively investigate the near extremal limit, with a result complementary to our perturbative investigation of the slow rotation limit~\cite{Baines:2024}. 
(For background, see~\cite{Lense:1918, Mashhoon:1984, Baines:2020, Baines:2021a, Baines:2021b, Baines:2022}.)
We plan to follow up and expand on the current results in a longer publication currently in preparation~\cite{Baines:2024-in-preparation}.

\vfill
\emph{Acknowledments:}
JB was supported by  a Victoria University of Wellington PhD Doctoral Scholarship.
During early phases of this work MV was supported by the Marsden Fund, via a grant administered by the Royal Society of New Zealand.

\clearpage

\end{document}